\documentclass[a4paper,fleqn]{cas-sc}

\usepackage[authoryear,longnamesfirst]{natbib}
\usepackage{pdflscape}
\usepackage{tabularx}
\usepackage{float}

\def\tsc#1{\csdef{#1}{\textsc{\lowercase{#1}}\xspace}}
\tsc{WGM}
\tsc{QE}

\begin{document}
\let\WriteBookmarks\relax
\def\floatpagepagefraction{1}
\def\textpagefraction{.001}

\shorttitle{}    

\shortauthors{}  

\title [mode = title]{Uncovering Local Heterogeneity: Local Summary Characteristics for Spatial Point Processes with Composition-Valued Marks}  



%

\author[1]{Clemens Baldzuhn}[orcid=0009-0008-1023-865X]



\ead{clemens.baldzuhn@gesis.org}



\affiliation[1]{organization={Department Computational Social Science, GESIS – Leibniz Institute for the Social Sciences}, 
            city={Cologne},
            country={Germany}}

\author[2]{Matthias Eckardt}[orcid=0000-0002-4015-486X]


\cormark[1] 

\ead{m.eckardt@hu-berlin.de}



\affiliation[2]{organization={Chair of Statistics, Humboldt-Universität zu Berlin},
            city={Berlin},
            country={Germany}}

\cortext[1]{Corresponding author}



\begin{abstract}
Traditional analysis of marked spatial point processes often relies on global summary statistics, which tend to obscure local spatial heterogeneity by averaging dependencies across the entire observation window. To overcome this limitation, this paper introduces a framework for Local Indicators of Mark Association (LIMA) specifically designed for composition-valued marks. Such marks, characterized by their non-negative components and sum-to-constant constraint, require a specialized treatment within the Aitchison geometry. By employing log-ratio transformations, we project these constrained marks into a Euclidean space, enabling the point-specific decomposition of global mark characteristics. The efficacy of the proposed clr-based LIMA functions is validated through extensive simulation studies. The results demonstrate a superior capacity to detect localized mark clusters, achieving detection accuracies consistently higher than their global counterparts.  

The practical utility of this framework is demonstrated using an empirical dataset of economic sector compositions in Castile-La Mancha, Spain. The analysis uncovers latent economic clustering patterns and localized \textit{drainage} effects that are invisible to global metrics, providing granular insights into regional spatial dynamics. Our findings suggest that the extended LIMA framework serves as a vital diagnostic tool for high-dimensional, non-stationary marked point patterns. 
\end{abstract}


\maketitle

\section{Introduction}\label{introduction}
Marked spatial point processes serve as a crucial framework for modeling spatial structures in which events are randomly situated and linked to additional attributes, termed \emph{marks}. In recent years, there has been a significant shift toward analyzing complex settings where marks are non-scalar quantities residing in non-Euclidean spaces. Typical examples include function-valued marks \citep{Comas2011, Eckardt2023MultiFunctionMarks}, graph-valued marks \citep{eckardt2024secondordercharacteristicsspatialpoint}, and \emph{composition-valued marks} \citep{eckardtSpatialPointProcesses}. The latter are particularly relevant in fields such as ecology, geology, and economics, where observations describe a whole quantity in terms of fractions of its components, such as employment shares across different economic sectors.

The distributional properties of these marks over space are traditionally investigated using functional summary characteristics, such as the mark correlation function or the mark variogram. These metrics describe the average statistical behavior of marks relative to the spatial distance $r$ between points \citep{Eckardt:Moradi:currrent}. However, a notable limitation of these established tools is their reliance on a global perspective. By calculating spatial averages across the entire observation area, traditional summary characteristics capture only the overall mark dependence structure. This inherent averaging process is susceptible to masking local heterogeneity; strong clusters of highly dependent marks in one sub-area may be overshadowed by weak dependence elsewhere, or outliers may create a false impression of global correlation.

The conceptual solution to this problem lies in the shift from global to local indicators. In the analysis of lattice data and regional aggregates, this shift was pioneered by the development of \emph{Local Indicators of Spatial Association} (LISA), introduced by \citet{anselinLocalIndicatorsSpatial}. LISA statistics, such as local Moran's $I$, allow for the decomposition of global measures into the contribution of each individual observation, enabling the detection of local "hotspots", "coldspots", and spatial outliers.

While LISA has become a standard tool for area-based data, its application to marked point processes is not straightforward due to the continuous nature of space and the specific dependence between point locations and their marks. To address this, \citet{eckardtLocalIndicatorsMark} bridged this gap by introducing \emph{Local Indicators of Mark Association} (LIMA) for real-valued and function-valued marks. Providing a point-specific decomposition of global mark second-order characteristics, LIMA functions represent the point-process analogue to LISA that allow researchers to evaluate the mark distribution around a fixed point $x_i$ relative to its neighbors.

The main advancement of this paper is to extend the LIMA framework to the domain of compositional data analysis (CoDa). Since the simplex, i.e. the natural space for compositional marks, possesses a non-standard geometry, standard statistical operations are inapplicable \citep{aitchisonStatisticalAnalysisCompositional}. 
Building on the work of \citet{eckardtSpatialPointProcesses}, we utilize log-ratio transformations to project composition-valued marks into a Euclidean space where LIMA functions can be formally defined and interpreted. This allows for the detection of localized associations or variations within specific compositional parts (componentwise) or the entire composition (compositional) that would remain hidden in a global analysis. By identifying whether a specific municipality or biological sample acts as a local catalyst for mark similarity, this approach provides a granular diagnostic capability that global measures inherently lack.

The paper is structured as follows. Section~\ref{ch:background} provides the mathematical preliminaries for marked spatial point processes and the construction principles for summary characteristics for real-valued marks. Moreover, details on the geometric properties of the simplex and the necessary log-ratio transformation methods are given. Section~\ref{ch:codasumstats} then introduces the formal definition of local summary characteristics for composition-valued marks. To evaluate the statistical power and type I error rates of these new metrics, we present a comprehensive simulation study in section ~\ref{ch:simulation}. Finally, Section~\ref{ch:application} demonstrates the practical efficacy of the proposed methods by analyzing an empirical dataset of economic sector compositions in the Spanish region of Castile-La Mancha.

\section{Background}
\label{ch:background}

\subsection{Marked spatial point processes}
Let $X = \{ (x_i, m_i)\}_{i= 1}^{n}$ be a simple marked spatial point process defined on the product space $\mathbb{R}^d \times \mathbb{M}$, where the ground process lies in $\mathbb{R}^d$ and marks belong to a Polish space $\mathbb{M}$. We assume $X$ to be motion-invariant (stationary and isotropic), implying a constant intensity $\lambda$. The first-order moment measure is given by $\Lambda(B \times L) = \lambda |B| M(L)$, where $|B|$ is the Lebesgue measure of a spatial subset $B$ and $M(L)$ denotes the mark distribution with probability density $f_M(m)$. We denote the mean mark by $\mu$ and the mark variance by $\sigma^2$.

To characterize the dependence structure of $X$, we employ summary characteristics based on the second-order product density $\varrho^{(2)}((x_1, m_1), (x_2, m_2))$. Under the assumption of motion-invariance, this density depends only on the inter-point distance $r = \|x_1 - x_2\|$ and factorizes into $\varrho^{(2)}(r, m_1, m_2) = \varrho^{(2)}(r) M_r(m_1, m_2)$, where $\varrho^{(2)}(r)$ is the product density of the ground process and $M_r$ is the two-point mark distribution.
Mark summary characteristics evaluate the expectation of a test function $\mathbf{t}_f: \mathbb{M} \times \mathbb{M} \rightarrow \mathbb{R}^+$ conditional on the presence of points at distance $r$. Following \citet{schlatherDetectingDependenceMarks} and \citet{illianStatisticalAnalysisModelling}, the unnormalized characteristic is defined as
\begin{equation*}
\label{density_frac2}
    \nabla_{\mathbf{t}_f}(r) = \frac{\varrho_{\mathbf{t}_f}^{(2)}(r)}{\varrho^{(2)}(r)} = \int_{\mathbb{M}}\int_{\mathbb{M}}\mathbf{t}_f(m_1,m_2) M_r(\mathrm{d}(m_1,m_2)).
\end{equation*}
The normalized $\mathbf{t}_f$-correlation function is given by $\kappa_{\mathbf{t}_f}(r) = \nabla_{\mathbf{t}_f}(r) / \nabla_{\mathbf{t}_f}(\infty)$, where the denominator represents the expectation under independent marks.

We consider three classical global summary characteristics for real-valued marks ($\mathbb{M} = \mathbb{R}$) where $m_1$ is the mark at the origin $\circ$ and $m_2$ the mark at a point at an interpoint distance $\Vert\mathbf{r}\Vert$=r.
1. The \textit{mark correlation function} $\kappa(r)$ with $\mathbf{t}_1(m_1, m_2) = m_1 m_2$, normalized by $\mu^2$,
2. the \textit{mark variogram} $\gamma(r)$ with $\mathbf{t}_2(m_1, m_2) = 0.5(m_1 - m_2)^2$, approaching $\sigma^2$ for $r \to \infty$, and
3. \textit{Shimatani's I} with $\mathbf{t}_3(m_1, m_2) = (m_1 - \mu)(m_2 - \mu)$, analogous to Moran's I \citep{shimataniPointProcessesFineScale}.

Local counterparts of these characteristics, denoted by index $i$, describe the relationship from the perspective of a fixed point $(x_i, m_i)$ via Palm-conditional expectations $\mathbb{E}_i[\mathbf{t}_{f,i}(m_i, m_k)]$ with $m_k$ denoting the mark at any alternative location $i \neq k$ at an interpoint distance $r$ \citep{eckardtLocalIndicatorsMark}.

\subsection{Compositional data and log-ratio transformations}

In this work, the mark space corresponds to the $D$-part simplex $\mathbb{S}^D = \{ \mathbf{c} \in \mathbb{R}^D \mid c_j > 0, \sum c_j = \mathfrak{K} \}$. Statistical analysis on $\mathbb{S}^D$ relies on Aitchison geometry to account for the constant-sum constraint and the relative nature of the data \citep{aitchisonStatisticalAnalysisCompositional}.
Instead of operating directly on the simplex, we apply the principle of working in coordinates \citep{mateu-figuerasPrincipleWorkingCoordinates}, mapping compositions to Euclidean space via log-ratio transformations $\psi: \mathbb{S}^D \to \mathbb{R}^{\tilde{D}}$.

The \textit{additive log-ratio transformation} (alr) maps $\mathbf{c}$ to $\mathbb{R}^{D-1}$ via $\psi_j(\mathbf{c})=\log(c_j/c_D)$ for $j=1,\dots,D-1$. While isomorphic, it is not isometric and treats components asymmetrically.
The \textit{centered log-ratio transformation} (clr) preserves symmetry and Euclidean distances:
\begin{equation*}
    \psi_j(\mathbf{c})=\operatorname{clr}_j(\mathbf{c}) = \log(c_j/g(\mathbf{c})), \quad j=1,\dots,D,
\end{equation*}
where $g(\mathbf{c})$ is the geometric mean. The resulting vector lies in a subspace of $\mathbb{R}^D$ summing to zero, leading to singular covariance matrices.
The \textit{isometric log-ratio transformation} (ilr) provides an isometric mapping to $\mathbb{R}^{D-1}$ by using an orthonormal basis $\{e^*_1, \dots, e^*_{D-1}\}$:
\begin{equation*}
    \operatorname{ilr}(\mathbf{c}) = \left( \langle \operatorname{clr}(\mathbf{c}), e^*_1 \rangle, \dots, \langle \operatorname{clr}(\mathbf{c}), e^*_{D-1} \rangle \right).
\end{equation*}
Throughout this paper, we assume all compositional parts are strictly positive to ensure these transformations are well-defined.

\section{Summary characteristics for composition-valued marks}
\label{ch:codasumstats}

Building on the established framework for compositional data, we define summary characteristics for marked spatial point processes with composition-valued marks $m_i = \mathbf{c}(x_i)$. We distinguish between global characteristics, following \citet{eckardtSpatialPointProcesses}, and local characteristics, which constitute our main contribution.
We denote the $j$-th part of a composition as $m_i^j$. Mark summary characteristics operate in Euclidean space on $\psi$-transformed compositions. A test function operating on the $j$-th and $l$-th parts of $\psi$-transformed marks $m_1$ and $m_2$ is denoted by $\mathbf{t}^{\psi,jl}_f(m_1^{\psi,j},m_2^{\psi,l})$.
While \citet{eckardtSpatialPointProcesses} also considers operations on complete compositions via Aitchison geometry, we restrict our presentation to component-wise formulations which investigate the association/variation between distinct parts of a $D$-part composition-valued mark.

\subsection{Global summary characteristics}
The global mark summary characteristic is defined as the expectation of a test function conditional on the presence of points with distance $r$:
\begin{equation*}
\label{global_cond_sumstat}
\nabla_{\mathbf{t}_f}^{\psi,jl}(r) = \mathbb{E}_{\circ}\left[\mathbf{t}_f^{\psi,jl}(m_1^{\psi,j},m_2^{\psi,l})\right].
\end{equation*}
Setting $j \neq l$ yields a cross-type characteristic, while $j=l$ describes the auto-relation of a component. Under independence, i.e. when the distance $r$ tends to infinity, the characteristic converges to
\begin{equation*}
    \label{tf_infty_glob}
    \nabla_{\mathbf{t}_f}^{\psi,jl}(\infty) = \int_{\mathbb{R}^{\tilde{D}}}\int_{\mathbb{R}^{\tilde{D}}}\mathbf{t}_f^{\psi,jl}(m_1^{\psi,j},m_2^{\psi,l})M(\operatorname{d}m_\circ^{\psi,j})M(\operatorname{d}m_k^{\psi,l}).
\end{equation*}
The normalized $\mathbf{t}_{f}^{\psi,jl}$-correlation function is then given by $\kappa_{\mathbf{t}_f}^{\psi,jl}(r) = {\nabla_{\mathbf{t}_f}^{\psi,jl}(r)}/{\nabla_{\mathbf{t}_f}^{\psi,jl}(\infty)}$.
We focus on three specific characteristics: the mark correlation function ($\mathbf{t}_1$), mark variogram ($\mathbf{t}_2$), and Shimatani's $I$ ($\mathbf{t}_3$).

For instance, employing the clr-transformation ($\psi = \text{clr}$) and the test function $\mathbf{t}_{1}^{\mathrm{clr},jl} = m_1^{\mathrm{clr},j}m_2^{\mathrm{clr},l}$, the conditional mean product of marks is
\begin{align*}
\tau^{\mathrm{clr},jl}(r) = \mathbb{E}_{\circ}\left[\log\left(\frac{m_1^{j}}{g(\mathbf{c})}\right)\cdot\log\left(\frac{m_2^{l}}{g(\mathbf{c})}\right)\right],
\end{align*}
where $g(\mathbf{c})$ is the geometric mean. The corresponding normalization factor is $\nabla_{\mathbf{t}_1}^{\psi, jl}(\infty) = \mu^{\psi, j} \mu^{\psi, l}$, reducing to $(\mu^{\psi,j})^2$ for auto-relations ($j=l$). Analogous derivations apply to the mark variogram and Shimatani's $I$ \citep{eckardtSpatialPointProcesses}.

\subsection{Local summary characteristics}
\begin{landscape}
\begin{table}[ht]
\caption{Summary of the notation used to denote test functions for creating mark summary characteristics with different scopes. Test functions $\mathbf{t}_1$ construct mark correlation functions, $\mathbf{t}_2$ mark variograms, and $\mathbf{t}_3$ Shimatani's $I$.}
\scriptsize
\centering
\begin{tabularx}{\linewidth}{lllll X}
\toprule
\makecell[l]{\textbf{Name}} & 
\makecell[l]{\textbf{Test} \\ \textbf{function} $\mathbf{t}_{f,i}^{\psi,jl}$} & 
\makecell[l]{\textbf{Normalizing} \\ \textbf{factor} $\nabla_{\mathbf{t}_{f,i}}^{\psi,jl}(\infty)$} & 
\makecell[l]{\textbf{Notation} \\ \textbf{for} $\nabla_{\mathbf{t}_{f,i}}^{\psi,jl}(r)$} & 
\makecell[l]{\textbf{Notation} \\ \textbf{for} $\kappa_{\mathbf{t}_{f,i}}^{\psi,jl}(r)$}  &
\makecell[l]{\textbf{Interpretation}} \\
\midrule
\multicolumn{6}{l}{\textbf{1. Local Summary Characteristics for Real Valued Marks}}\\
\addlinespace
$\mathbf{t}_{1,i}$ & $m_i m_k$ & $m_i\mu$ & $\tau_i(r)$ & $\kappa_i(r)$ & Local association: Degree of resemblance (joint magnitude) between $m_i$ and its neighbor $m_k$\\
$\mathbf{t}_{2,i}$ & $0.5(m_i - m_k)^{2}$ & $\zeta_i$ & $\gamma_i(r)$ & $\Gamma_i(r)$ & Local variability: Degree of dissimilarity (squared difference) between $m_i$ and $m_k$\\
$\mathbf{t}_{3,i}$ & $m_i(m_k-\mu)$ & - & $\iota_i(r)$ & - & Local correlation: Tendency for neighbor $m_k$ of $m_i$ to be above (attraction of marks) or below (repulsion of marks) the population average.\\
\midrule
\multicolumn{6}{l}{\textbf{2. Global Summary Characteristics for Composition-Valued Marks}}\\
\addlinespace
$\mathbf{t}_{1}^{\psi,jl}$ & $m_1^{\psi,j}m_2^{\psi,l}$ & $\mu^{\psi, j}\mu^{\psi,l}$ & $\tau^{\psi,jl}(r)$ & $\kappa^{\psi,jl}(r)$ & Average association: Global degree of resemblance between component $j$ and component $l$ in the neighborhood. \\
$\mathbf{t}_{2}^{\psi,jl}$ & $0.5(m_1^{\psi,j} - m_2^{\psi,l})^{2}$ & $\zeta^{\psi,jl}$ & $\gamma^{\psi,jl}(r)$ & $\Gamma^{\psi,jl}(r)$ & Average variability: Global degree of dissimilarity between component $j$ and component $l$ in the neighborhood. \\
$\mathbf{t}_{3}^{\psi,jl}$ & $(m_1^{\psi,j}-\mu^{\psi,j})( m_2^{\psi,l}-\mu^{\psi,l})$ & $\sigma^{\psi,jl}$& $\iota^{\psi,jl}(r)$ & $I^{\psi,jl}(r)$ & Average correlation: Global tendency for deviations in component $j$ to coincide with deviations in $l$ (attraction/repulsion of component types). \\\midrule
\multicolumn{6}{l}{\textbf{3. Local Summary Characteristics for Composition-Valued Marks}}\\
\addlinespace
$\mathbf{t}_{1,i}^{\psi,jl}$ & $m_i^{\psi,j} m_k^{\psi,l}$ & $m_i^{\psi,j} \mu^{\psi,l}$ & $\tau_{i}^{\psi,jl}(r)$  & $\kappa_i^{\psi,jl}(r)$  & Local association: Degree of resemblance between component $j$ at point $i$ and component $l$ at neighbor $k$. \\
$\mathbf{t}_{2,i}^{\psi,jl}$ & $0.5(m_i^{\psi,j} - m_k^{\psi,l})^{2}$ & $\zeta_i^{\psi,jl}$ & $\gamma_{i}^{\psi,jl}(r)$ & $\Gamma_i^{\psi,jl}(r)$  & Local variability: Degree of dissimilarity between component $j$ at point $i$ and component $l$ at neighbor $k$. \\
$\mathbf{t}_{3,i}^{\psi,jl}$ & $m_i^{\psi,j}(m_k^{\psi,l} - \mu^{\psi,l})$ & - & $\iota_i^{\psi,jl}(r)$ & -  & Local correlation: Tendency for neighbor $m_k^l$ to be above (attraction) or below (repulsion) the average, given the value of $m_i^j$. \\
\bottomrule
\end{tabularx}
\label{tb:loc_summstats}
\end{table}    
\end{landscape}
To evaluate local deviations from global behavior, we introduce local summary characteristics for composition-valued marks, bridging \citet{eckardtLocalIndicatorsMark} and \citet{eckardtSpatialPointProcesses}. By conditioning on a fixed point $(x_i, m_i)$, we define the local summary characteristic and its independence counterpart as:
\begin{align}
    \nabla_{\mathbf{t}_{f,i}}^{\psi,jl}(r) &= \mathbb{E}_{i}\left[\mathbf{t}_{f,i}^{\psi,jl}(m_{i}^{\psi,j}, m_{k}^{\psi,l})\right], \notag \\
    \label{tf_infty_loc}
    \nabla_{\mathbf{t}_{f,i}}^{\psi,jl}(\infty) &= \int_{\mathbb{R}^{\tilde{D}}}\int_{\mathbb{R}^{\tilde{D}}}\mathbf{t}_{f,i}^{\psi,jl}(m_i^{\psi,j}, m_k^{\psi,l}) M(\operatorname{d}m_i^{\psi,j}) M(\operatorname{d}m_k^{\psi,l}).
\end{align}
We propose three specific local characteristics: the local mark correlation function $\kappa_i^{\psi,jl}(r)$, the local mark variogram $\gamma_i^{\psi,jl}(r)$, and the local Shimatani's $I$ $\iota_{i}^{\psi,jl}(r)$. 

\paragraph{Local mark correlation function}
Using test function $\mathbf{t}_{1,i}^{\psi,jl} = m_i^{\psi,j} m_k^{\psi,l}$, the local conditional mean product for clr-transformed marks is
\begin{align*}
    \tau_{i}^{\mathrm{clr},jl}(r) = \mathbb{E}_{i}\left[m_i^{\mathrm{clr},j} m_k^{\mathrm{clr},l}\right].
\end{align*}
The normalization factor is derived as $\nabla_{{\mathbf{t}_{1,i}}}^{{\psi,jl}}(\infty) = m_{i}^{\psi,j} \mu^{\psi,l}$. The normalized function $\kappa_i^{\psi,jl}(r)$ equals 1 under random labeling, indicating indistinguishability from a random distribution of marks.

\paragraph{Local mark variogram}
Defined by $\mathbf{t}_{2,i}^{\psi,jl} = 0.5(m_i^{\psi,j} - m_k^{\psi,l})^{2}$, the local mark variogram is $\gamma_{i}^{\psi,jl}(r) = \mathbb{E}_{i}[\mathbf{t}_{2,i}^{\psi,jl}]$. The normalization factor is obtained by inserting the test function into (\ref{tf_infty_loc}):
\begin{align*}
    \zeta_i^{\psi, jl} = 0.5\left[ (m_i^{\psi,j}- \mu^{\psi,l})^2 + \sigma^{\psi,l} \right],
\end{align*}
where $\sigma^{\psi,l}$ is the variance of the $l$-th transformed part.

\paragraph{Local Shimatani's $I$}
Using $\mathbf{t}_{3,i}^{\psi,jl} = m_i^{\psi,j}(m_k^{\psi,l} - \mu^{\psi,l})$, we define $\iota_{i}^{\psi,jl}(r) = \mathbb{E}_{i}[\mathbf{t}_{3,i}^{\psi,jl}]$. No normalized version exists, as the expectation under random labeling is zero.

\subsection{Estimation}
Estimators for these local characteristics follow the ratio of second-order product densities \citep{illianStatisticalAnalysisModelling, eckardtSpatialPointProcesses}:
\begin{equation*}
    \widehat{\nabla_{\mathbf{t}_{f,i}}^{\psi,jl}}(r) = \frac{\widehat{\varrho_{\mathbf{t}_{f,i}}^{(2),\psi,jl}(r)}}{\widehat{\varrho^{(2)}(r)}}.
\end{equation*}
The numerator is estimated via kernel smoothing:
\begin{equation*}
    \widehat{\varrho_{\mathbf{t}_{f,i}}^{(2),\psi,jl}(r)}=\frac{1}{2\pi r |W|}\sum_{(x_k,m_k)\in X}\mathbf{t}_{f,i}^{\psi,jl}\left( m_i^{\psi,j},m_k^{\psi,l}\right) \mathcal{K}\left( ||x_i-x_k||-r\right),
\end{equation*}
where $\mathcal{K}(\cdot)$ is a kernel function and $|W|$ the observation window area. The denominator $\widehat{\varrho^{(2)}(r)}$ is estimated analogously without the test function term.
The independence factor $\nabla_{\mathbf{t}_{f,i}}^{\psi,jl}(\infty)$ is estimated by the sample mean of the test function, $\frac{1}{n-1}\sum_{k}\mathbf{t}_{f,i}(m_i^{\psi,j},m_k^{\psi,l})$. We utilize the \textit{spatstat} library \citep{baddeleySpatstatSpatialPoint} for non-parametric density estimation.

\section{Simulation study}
\label{ch:simulation}

\begin{figure}
    \centering
    \includegraphics[width=\linewidth]{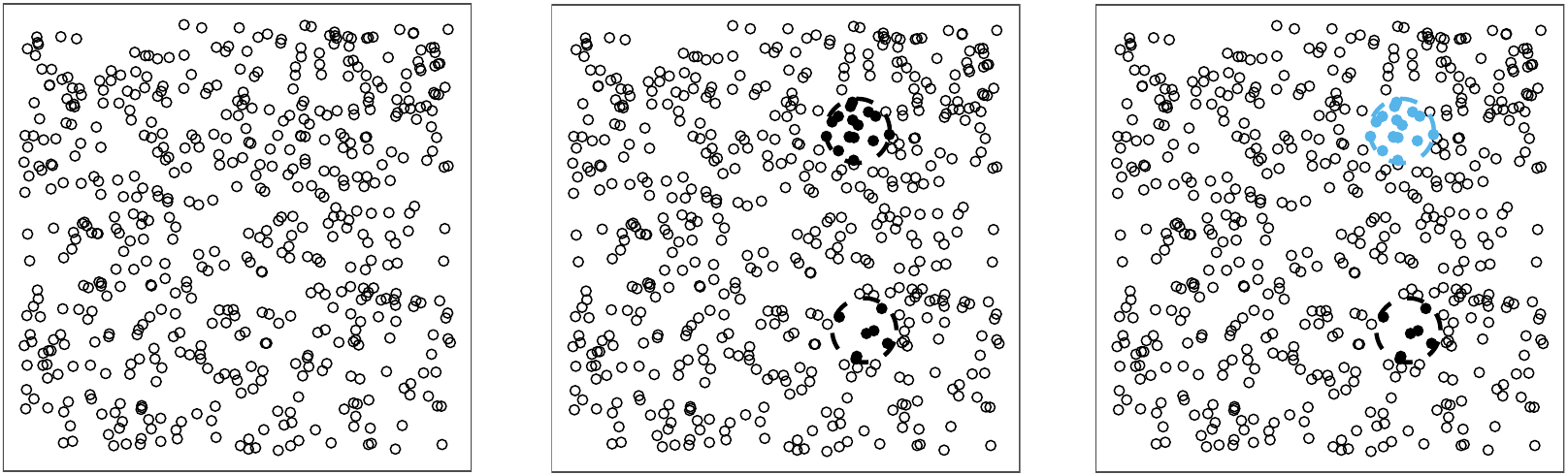}
    \caption{Visual representation of the dependence structures for scenarios I - III. To emphasize the spatial extent of the mark distributions, point locations are shown without compositional values. The specific Dirichlet-sampled compositions for these patterns are provided in the appendix (Fig. \ref{fig:appendix_eg_scenarios_samples}).}
    \label{fig:eg_scenarios}
\end{figure}

\begin{figure*}
    \centering
    \includegraphics[width=\linewidth]{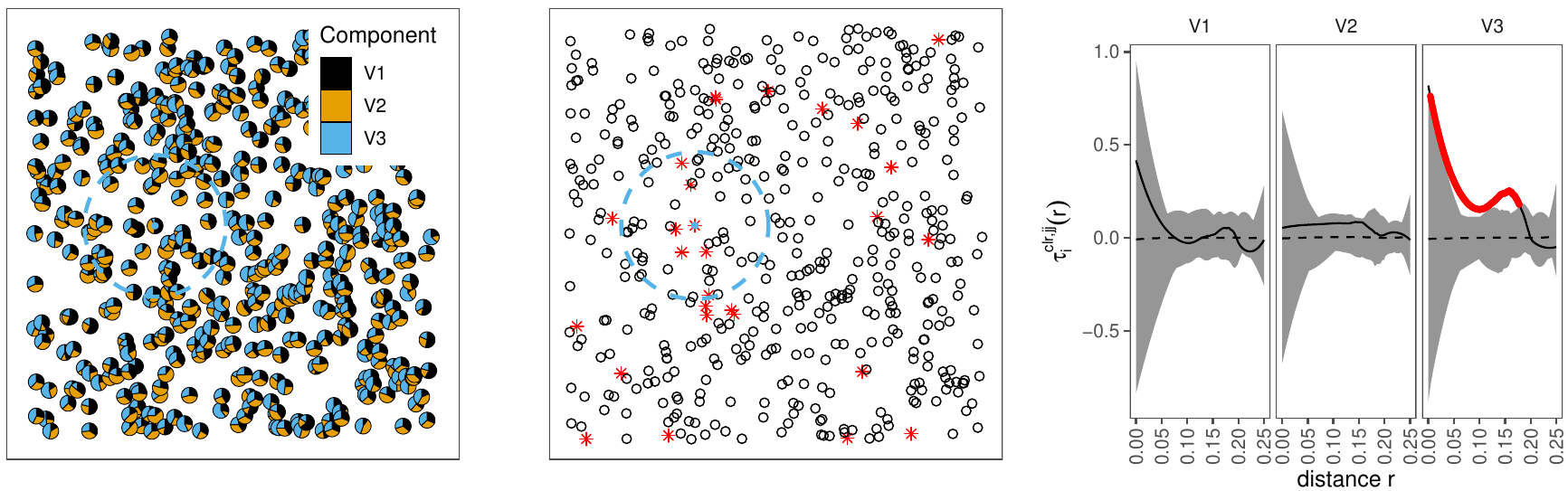}
    \caption{Diagnostic analysis of Scenario I. (Left) A pattern incorrectly flagged by the global characteristic. (Middle/Right) Local analysis reveals that the significance is driven by a small, localized cluster of points with high $V3$ associations, providing a granular view of the stochastic noise.}
    \label{fig:sim1}
\end{figure*}

To evaluate the performance and sensitivity of the proposed local summary characteristics, we conduct a controlled simulation study. We specifically compare the local conditional mean product of marks $\tau_i^{\mathrm{clr},jj}$ against its global counterpart $\tau^{\mathrm{clr},jj}$. Our primary hypotheses are twofold: (i) both measures should exhibit similar type I error rates under random labeling, and (ii) the local characteristics will demonstrate superior statistical power in identifying localized, non-stationary mark dependencies that are often smoothed out by global averaging.
  
We generated 500 spatial point patterns using a homogeneous Poisson process with intensity $\lambda = 500$ on a unit square $W = [0,1]^2$. To ensure that differences in the summary statistics are attributable solely to mark dependencies rather than fluctuations in the ground process, the same 500 point configurations were reused across all three scenarios. 

Compositional marks $\mathbf{c}(x_i) \in \mathbb{S}^3$ were sampled from Dirichlet distributions, $\text{Dir}(\alpha)$, where the parameter vector $\alpha$ controls the concentration and mean of the compositions (visualizations available in the appendix).

In Scenario I, we establish a baseline for null-hypothesis testing by sampling marks from a symmetric Dirichlet distribution ($\alpha_1 = (5,5,5)$) for all points, representing a state of complete spatial randomness in the marks. 
Scenario II introduces spatial heterogeneity via two discs of radius $r=0.075$, covering approximately 3.5\% of the window. Inside these discs, marks follow a $\text{Dir}(20,5,5)$ distribution, inducing a local positive association in the first component ($V1$). 
Scenario III increases complexity by assigning $\text{Dir}(20,5,5)$ to the first disc and $\text{Dir}(5,5,20)$ to the second, thereby creating two disparate regions of local dependence in $V1$ and $V3$, respectively.
 
Significance is assessed using global envelope tests based on the extreme rank length measure \citep{myllymakiGlobalEnvelopeTests}. This non-parametric approach allows for a joint assessment of the test function across a distance range $r \in [0, 0.25]$. For each point $x_i$ in the local case, and for the pattern in the global case, 500 permutations were used to generate $p$-values at a significance level of $\alpha = 0.05$. 

The local analysis is computationally intensive; with approximately 500 points per pattern and 500 patterns per scenario, the study requires 250,000 individual global envelope tests per scenario. These were executed on a high-performance computing cluster (2x AMD EPYC 7713, 64 threads), totaling roughly 150 core-hours of computation.
 
In the absence of true dependence, the global test correctly failed to reject the null hypothesis in 95\% of cases. The local tests yielded a nearly identical false-positive rate of 5.1\%. While the error rates are balanced, Figure \ref{fig:sim1} illustrates the diagnostic advantage of the local approach: when a global test is significant due to stochastic fluctuations, local characteristics can pinpoint the specific points and components (e.g., $V3$) responsible for the deviation, preventing misinterpretation of global trends.
 
\begin{figure*}
    \centering
    \includegraphics[width=0.95\linewidth]{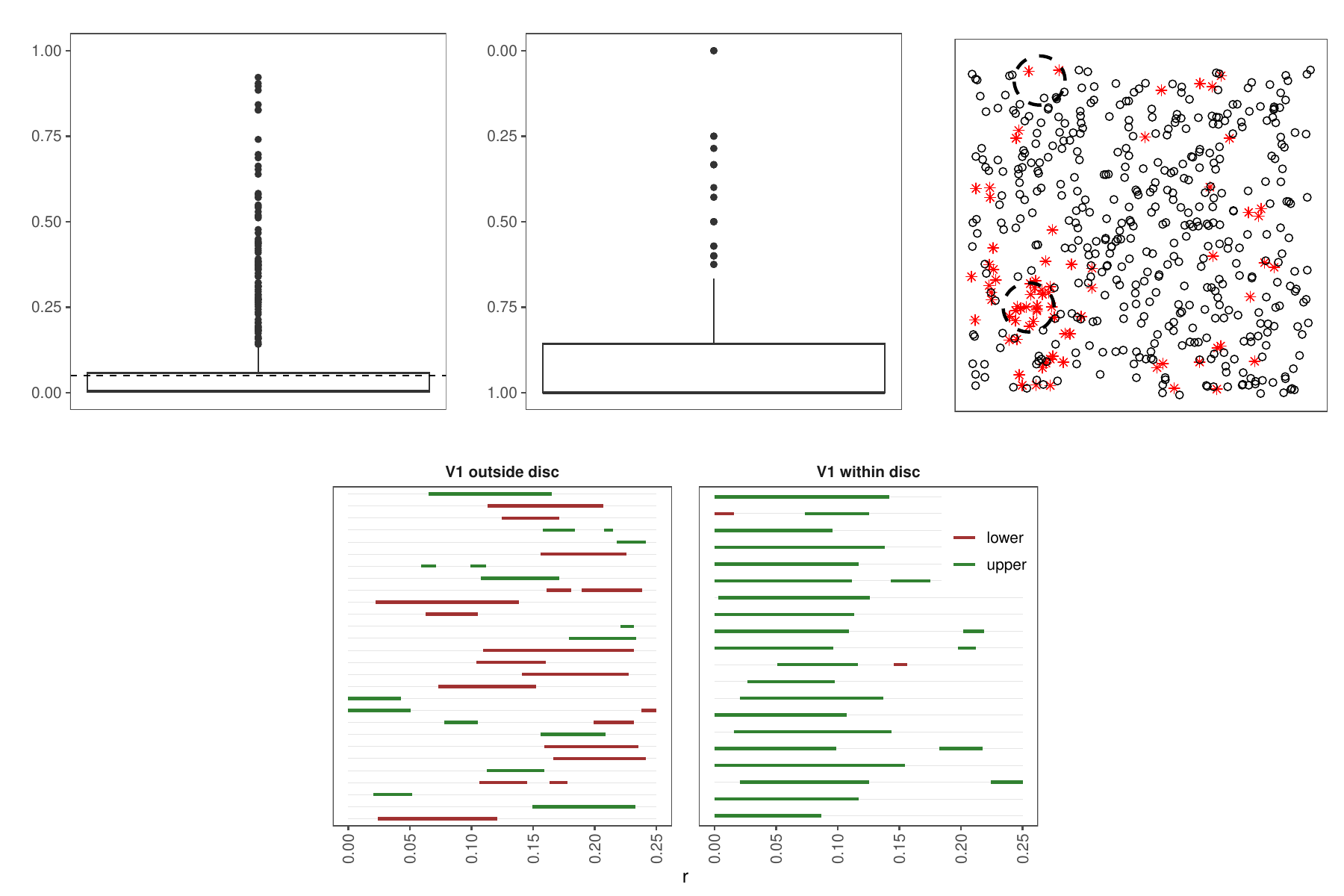}
    \caption{Results for Scenario II. The top row contrasts the global $p$-value distribution with the high local detection rate within the discs. The bottom panel displays the distance-dependent nature of these local associations, highlighting how points "sense" their environment.}
    \label{fig:sim2}
\end{figure*}

The results for the dependence scenarios confirm the limitations of global averaging. In Scenario II, the global characteristic detected mark dependence in 74.0\% of the patterns. In contrast, the local analysis correctly identified 88.9\% of the points situated within the dependent discs. As shown in Figure \ref{fig:sim2} (bottom), the local method effectively captures the "transition zone" at the disc boundaries, distinguishing between positive internal associations (green) and the negative associations (red) occurring relative to the background process.

For Scenario III, where two different components ($V1$ and $V3$) are non-stationary, the global power remained at 69.8\%. The local statistics, however, maintained a high detection rate of 91.9\% for the relevant points (Figure \ref{fig:sim3} in the appendix). This demonstrates that local characteristics are robust to "competing" dependence structures that might cancel each other out in global metrics.

In conclusion, the local conditional mean product of marks provides a powerful tool for detecting localized compositional dependencies. While global methods are sufficient for homogeneous processes, they lack the sensitivity required for complex, non-stationary patterns. Future studies should focus on the impact of point intensity and the selection of the smoothing kernel $\mathcal{K}$ on the resolution of these local indicators.

\newpage

\section{Application} \label{ch:application}
\begin{figure}
    \centering
    \includegraphics[width=0.5\linewidth]{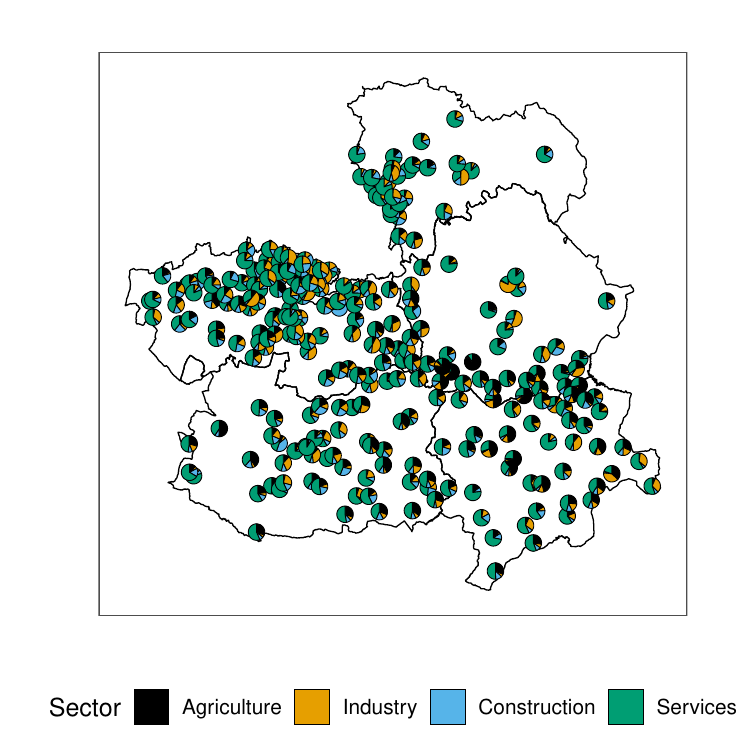}
    \caption{Economic sector compositions (2022) for 278 municipalities with $>1000$ inhabitants within the area of study Castile-La Mancha, a high plateau in central Spain.}
    \label{fig:dataplo}
\end{figure}

\begin{table}
\caption{Untransformed distributional statistics of the economic sector compositions.}
\centering
\footnotesize
\begin{tabular}{lrrrrlr}
\toprule
Sector & Min & Q1 & Mean & Median & Q3 & Max\\
\midrule
Agriculture & 0.1 & 4.9 & 16.0 & 12.0 & 22.5 & 85.3\\
Industry & 1.5 & 8.2 & 17.7 & 13.9 & 24.7 & 79.1\\
Construction & 1.7 & 8.0 & 12.7 & 11.3 & 16.3 & 59.1\\
Services & 10.7 & 44.2 & 53.7 & 53.0 & 62.0 & 91.0\\
\bottomrule
\end{tabular}
\label{tb:sum_stat_clm}
\end{table}

In this section, we apply the developed local summary characteristics to real-world data on economic sector compositions in Castile-La Mancha, Spain. This autonomous region, characterized by its high plateau and relatively homogeneous physiographical conditions, serves as an ideal study area for spatial point processes, as it minimizes the influence of extreme topographic variance on municipality locations \citep{ripleyModellingSpatialPatterns, diggleSpatialSpatioTemporalLogGaussian}.

The data comprises employment contracts categorized into four primary sectors: \textbf{Agriculture}, \textbf{Industry}, \textbf{Construction}, and \textbf{Services}. Preliminary statistics (Table \ref{tb:sum_stat_clm}) reveal a highly skewed distribution, dominated by the service sector (Mean: 53.7\%, Max: 91.0\%). This reflects the modern economic structure of central Spain, where administrative and commercial hubs drive employment. Conversely, the agricultural sector, while having a lower mean (16.0\%), shows the highest variability (Max: 85.3\%), suggesting specialized agrarian pockets within the region.

\subsection{Global analysis: Average spatial trends}
The global analysis serves to detect whether the spatial arrangement of sector proportions deviates from a random labeling scenario. Applying global envelope tests with $\tau^{\mathrm{clr},jj}$, $\gamma^{\mathrm{clr},jj}$, and $\iota^{\mathrm{clr},jj}$ yields significant results ($p < 0.02$) for all components.

\begin{figure}
    \centering
    \includegraphics[width=0.8\linewidth]{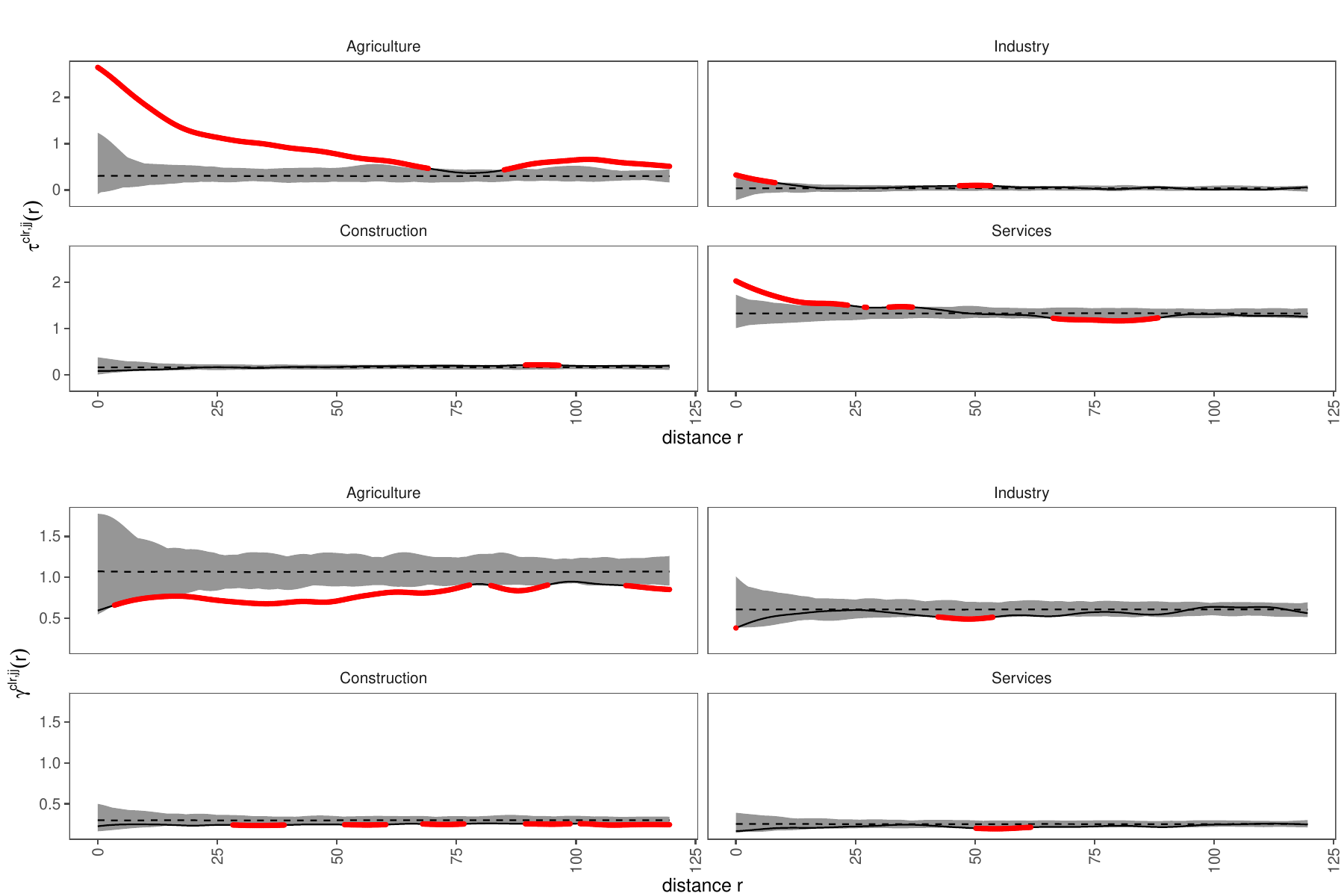}
    \caption{Global summary characteristics for clr-transformed marks. (Top) Conditional mean product $\tau^{\mathrm{clr},jj}$. (Bottom) Mark variogram $\gamma^{\mathrm{clr},jj}$. Observed values (black) outside the 95\% envelopes indicate significant global dependence.}
    \label{fig:app_globtests}
\end{figure}

As shown in Figure \ref{fig:app_globtests}, the agricultural sector exhibits a significant deviation from the null hypothesis across nearly all distances $r$. The fact that the empirical mark variogram $\gamma^{\mathrm{clr},jj}$ remains below the envelope indicates that municipalities close to each other tend to have more similar agricultural proportions than expected by chance. Similarly, the service sector shows significant association, particularly at short distances ($r \leq 30$) and long-range intervals ($r \in [60, 80]$). While these global findings provide a clear "rejection" of spatial randomness, they offer a "flat" perspective: they suggest a regional trend of clustering but fail to distinguish whether this is driven by a single dominant hub or multiple small-scale interactions.

\subsection{Local analysis: Pinpointing economic interactions}
The transition to local summary characteristics reveals a much more nuanced economic landscape. By computing $\tau_i^{\mathrm{clr}}$ for each point, we move from a regional average to a point-specific diagnostic. Figure \ref{fig:app_loctest} in the appendix identifies clusters of significant interaction that the global analysis obscured.

\begin{landscape}
\begin{figure}
    \centering
    \includegraphics[width=\linewidth]{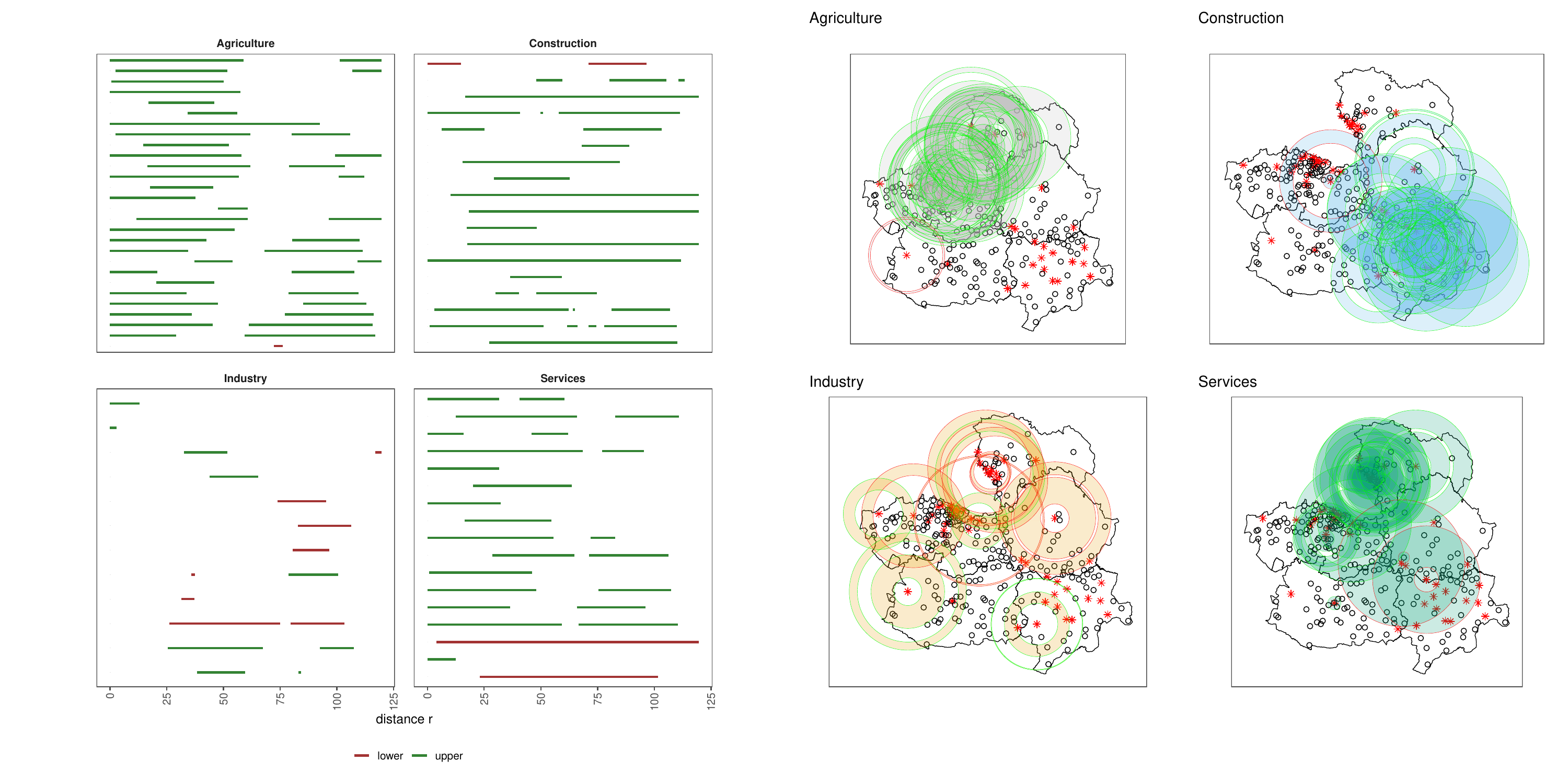}
    \caption{Spatial distribution of local significance bands ($\tau_i^{\mathrm{clr},jj}$). Green discs indicate significant local association; red discs indicate repulsion. These plots pinpoint where sectors behave similarly to their neighbors.}
    \label{fig:app_loctest_markcorr}
\end{figure}
\end{landscape}

The interpretation of the agricultural sector (cf. Fig. \ref{fig:app_loctest_markcorr}) is particularly revealing. While the global test suggested a region-wide dependency, the local analysis shows that these associations are heavily concentrated in the northwestern corridor near Madrid. In this area, the association actually stems from a shared absence or very low proportion of agricultural activity due to intense urbanization and proximity to the capital's service-driven economy. In contrast, the southwestern region shows associations driven by high agricultural density, representing the traditional "Don Quixote" plains where agriculture remains a primary employer.

For the \textbf{Industry} sector, the local analysis identifies a specific cluster south of Madrid (Guadalajara and Toledo provinces). Here, the autocorrelation (Shimatani's $I$, c.f. Fig. \ref{fig:appendix_loctest_shim}) indicates that industrial hubs are not isolated but form a contiguous "belt" of manufacturing. Interestingly, municipalities just outside this belt exhibit "repulsive" spatial dependence. Their industrial proportions are significantly lower than their neighbors', suggesting a "drainage" effect where industrial activity is sucked into the primary hubs.

The \textbf{Service} sector associations are most prominent in the north and northeast. While the raw data shows high service proportions everywhere, the local statistics highlight where these proportions are unusually high or low relative to the regional environment. The western border of Castile-La Mancha shows significant negative associations, pinpointing an economic "frontier" where the service sector is suppressed by a sudden increase in agricultural land use.

Finally, the \textbf{Construction} sector reveals long-distance associations in the southwest. This is a subtle finding that global metrics missed; it suggests that construction activity, which is often taken as a proxy for regional development and investment, moves in waves that affect groups of municipalities simultaneously, possibly linked to regional infrastructure projects or housing trends that span multiple administrative boundaries.

\section{Discussion and Conclusion}

This study has bridged the methodological gap between local indicators of spatial association and compositional mark analysis \citep{eckardtLocalIndicatorsMark, eckardtSpatialPointProcesses}. By extending construction principles to the local setting via log-ratio transformations, we provided a robust framework to analyze spatially varying mark dependencies within the Euclidean representation of the simplex.

Our simulation experiments and the empirical application to Castile-La Mancha confirm that local statistics exhibit superior sensitivity toward local clusters of mark dependence. Where global statistics tend to average out spatial heterogeneity, leading to a potentially misleading impression of regional uniformity, local characteristics pinpoint the exact "economic heartbeat" of the region. However, the study also highlights that local summary statistics are highly sensitive to the chosen distance range $r$ and the concentration of the compositional marks. The decrease in power observed in regular patterns during the simulation suggests that these metrics are best suited for "patchy" or non-stationary environments where local interactions dominate over global trends.

A key methodological consideration remains the choice between component-wise log-ratio analysis and direct compositional metrics in Aitchison geometry. While the log-ratio approach used here offers high interpretability for individual sectors (e.g., "Where is the agricultural sector clustering?"), it inherently sacrifices the "holistic" view of the composition. Future work could integrate these perspectives by developing local metrics that operate on the full simplicial distance between neighbor compositions. Furthermore, the current reliance on ground-process stationarity is a limitation. Generalizing these statistics to inhomogeneous settings via intensity-based spatial weighting \citep{moradiInhomogeneousMarkCorrelation} would allow researchers to account for the varying "density" of municipalities themselves, which often correlates with economic intensity.

In conclusion, the results manifest the significant advantages of local summary statistics for composition-valued marks. They offer the necessary sensitivity to detect small-scale dependencies and provide a granular understanding of regional economic, ecological, or social structures. By allowing researchers to visualize and quantify the "local sphere of influence" for each point, this methodology transforms marked point process theory into a precise diagnostic tool, capable of uncovering the complex, multi-dimensional dependencies inherent in modern spatial data.

\subsection*{Data and code availability}
Data and code to reproduce calculations and figures can be found on GitHub at \url{https://github.com/baldzuhnc/LIMACMarks}.

\newpage
\appendix \label{ch:appendix}

\begin{figure}[H]
    \centering
    \includegraphics[width=\linewidth]{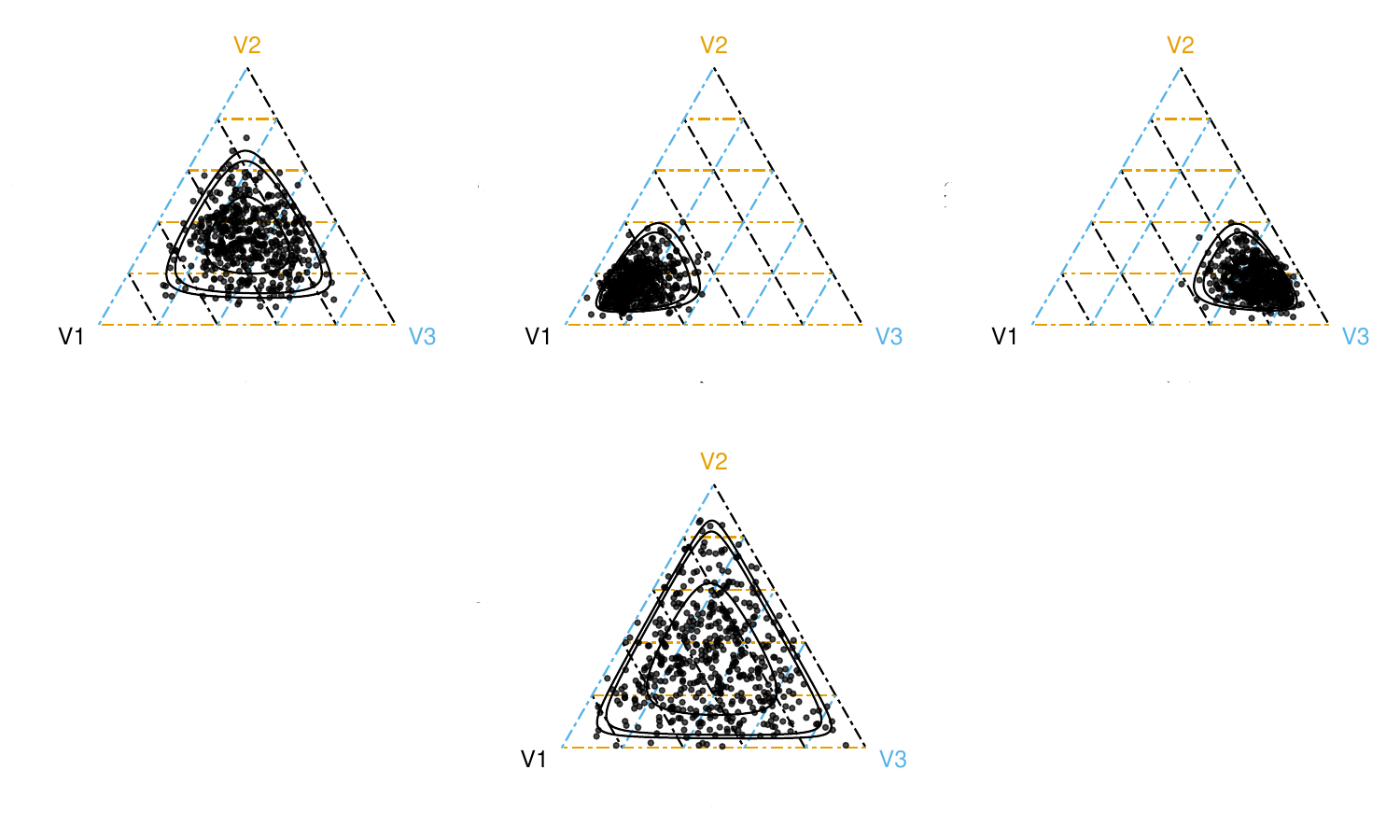}
    \caption{Visualization of 500 samples from a Dirichlet distribution with different parametrizations corresponding to those chosen for the simulation scenarios in chapter \ref{ch:simulation}. (Top row, left) A sample from a Dirichlet with $\alpha_1=(5,5,5)$. (Top row, middle): A sample with $\alpha_2=(20,5,5)$. (Top row, right) A sample with $\alpha_3=(5,5,20)$. Note that with alpha approaching one in each dimension, the distributional mass shifts to be more uniformly spread over the simplex and thereby increases the sample variance. This behavior is visualized in the bottom row figure which was sampled with $\alpha_4=(2,2,2)$.}
    
    \label{fig:appendix_eg_tern}
\end{figure}


\begin{figure*}
    \centering
    \includegraphics[width=\linewidth]{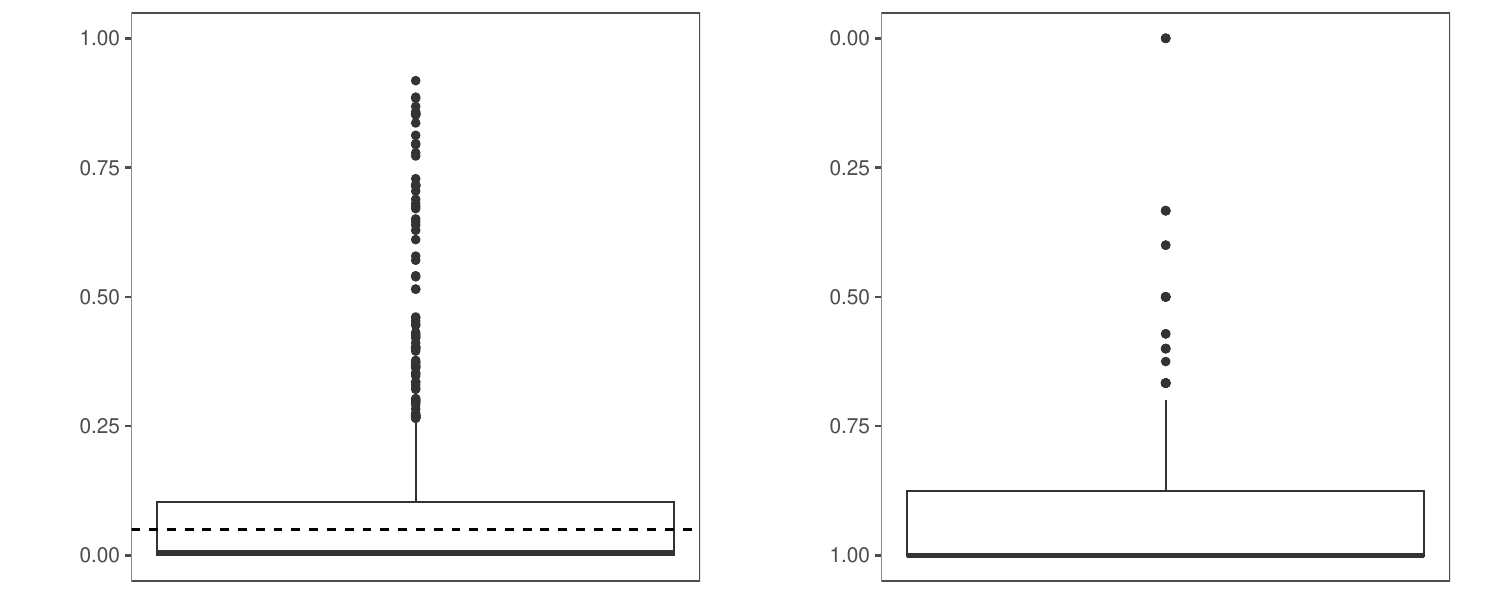}
    \caption{Comparison for Scenario III. The local approach (right) shows a significantly higher and more stable fraction of correctly detected dependent points compared to the global $p$-values (left), which exhibit higher variance.}
    \label{fig:sim3}
\end{figure*}

\begin{figure}[H]
    \centering
    \includegraphics[width=\linewidth]{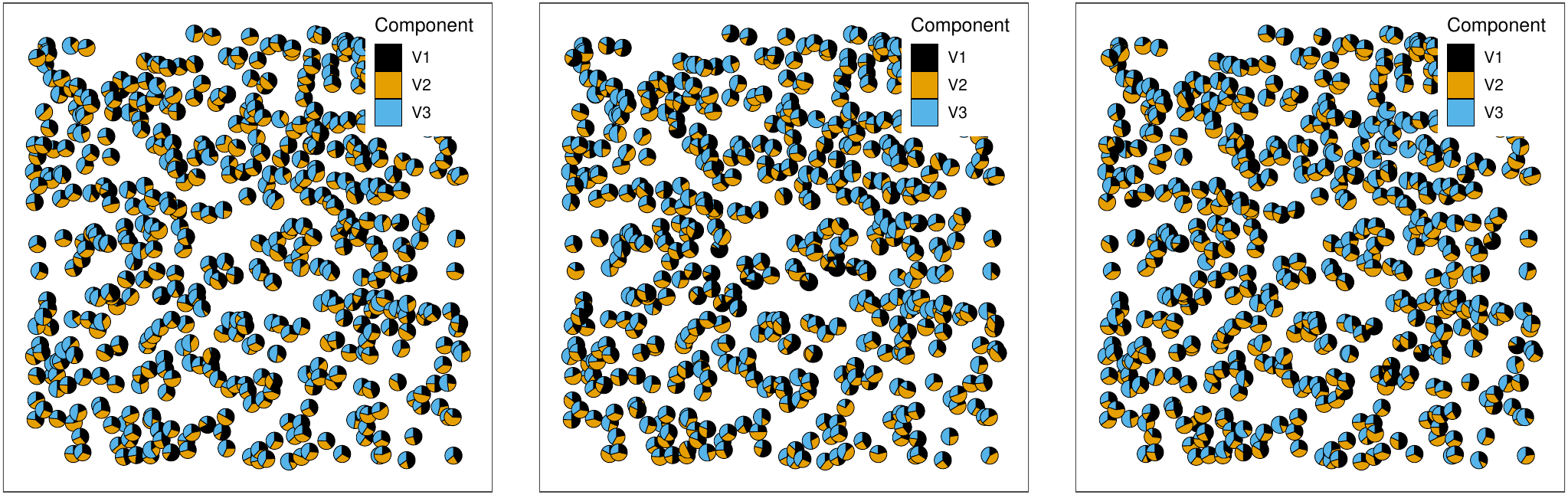}
    \caption{Examples for the mark dependence structures in the three simulation scenarios with sampled compositions. Patterns are the same as in Figure \ref{fig:eg_scenarios}.}
    \label{fig:appendix_eg_scenarios_samples}
\end{figure}

\begin{figure}[ht]
    \centering
    \includegraphics[width=\textwidth]{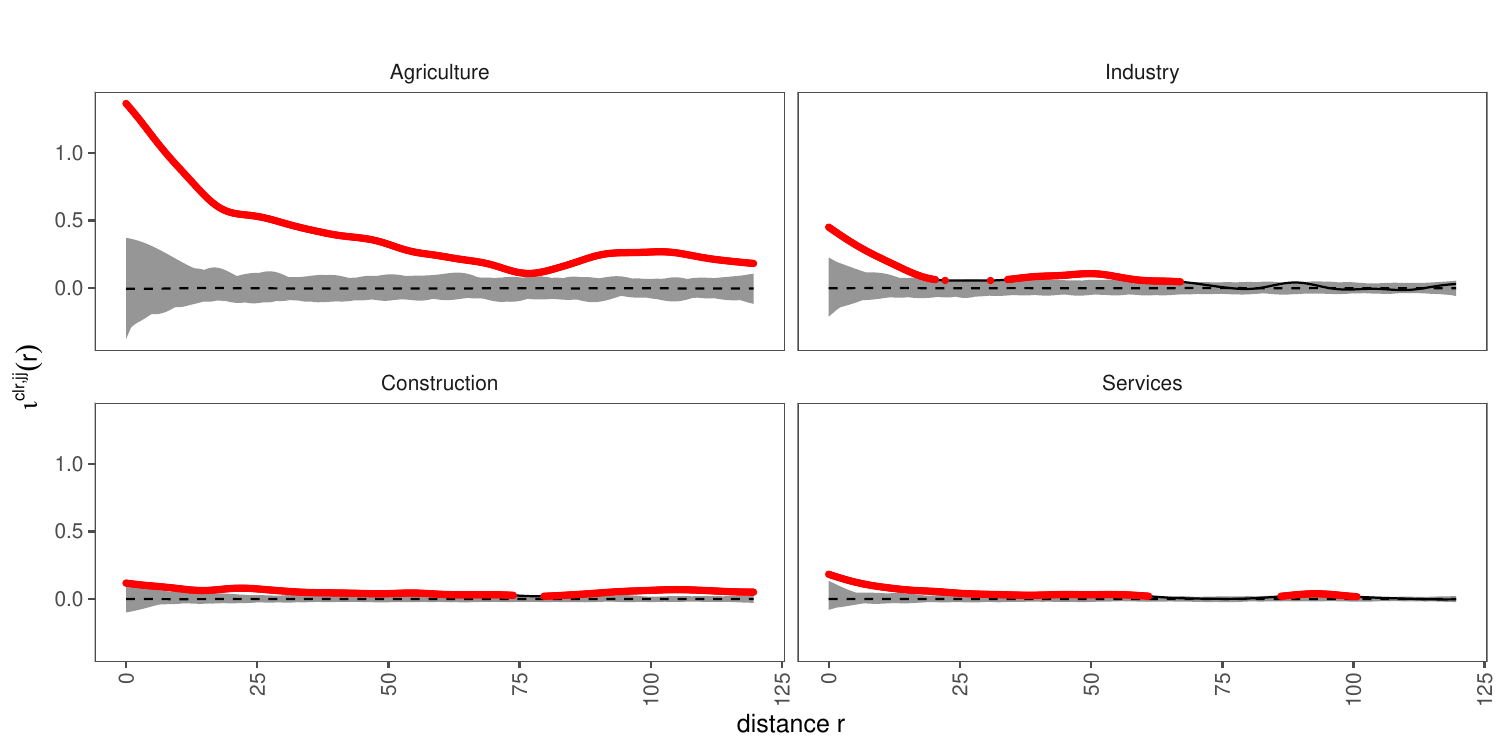}
    \caption{Results of the global test employing Shimatani's $I$.}
    \label{fig:appendix_glob_shim}
\end{figure}

\begin{figure}
    \centering
    \includegraphics[width=\linewidth]{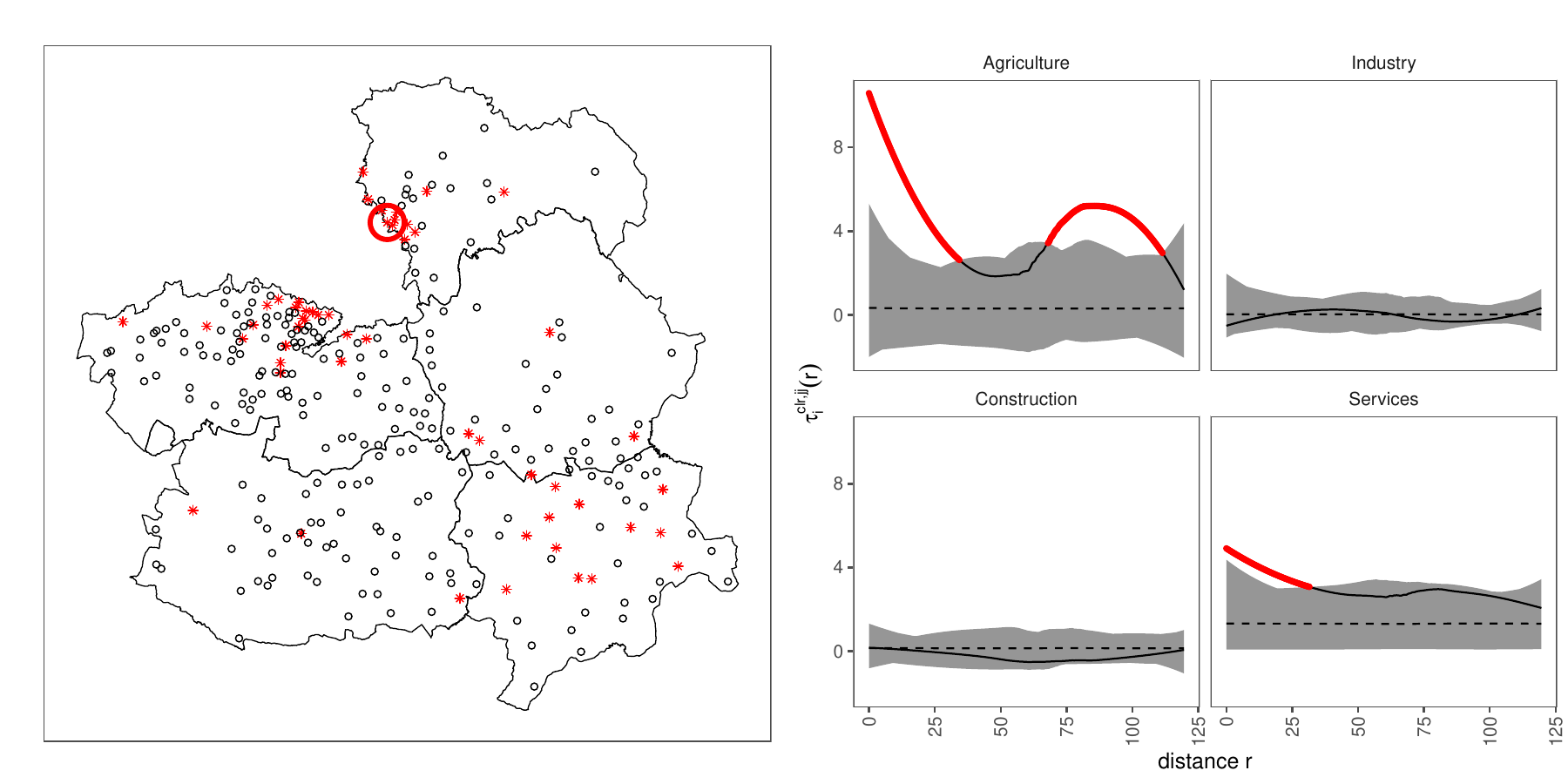}
    \caption{Local associations of sector proportions. (Left) Municipalities with significant global envelope tests (red stars). (Right) Point-specific results for a northern municipality, showing significant association in agriculture at short and long distances.}
    \label{fig:app_loctest}
\end{figure}

\begin{landscape}
\begin{figure}[h]
    \centering
    \includegraphics[width=\linewidth]{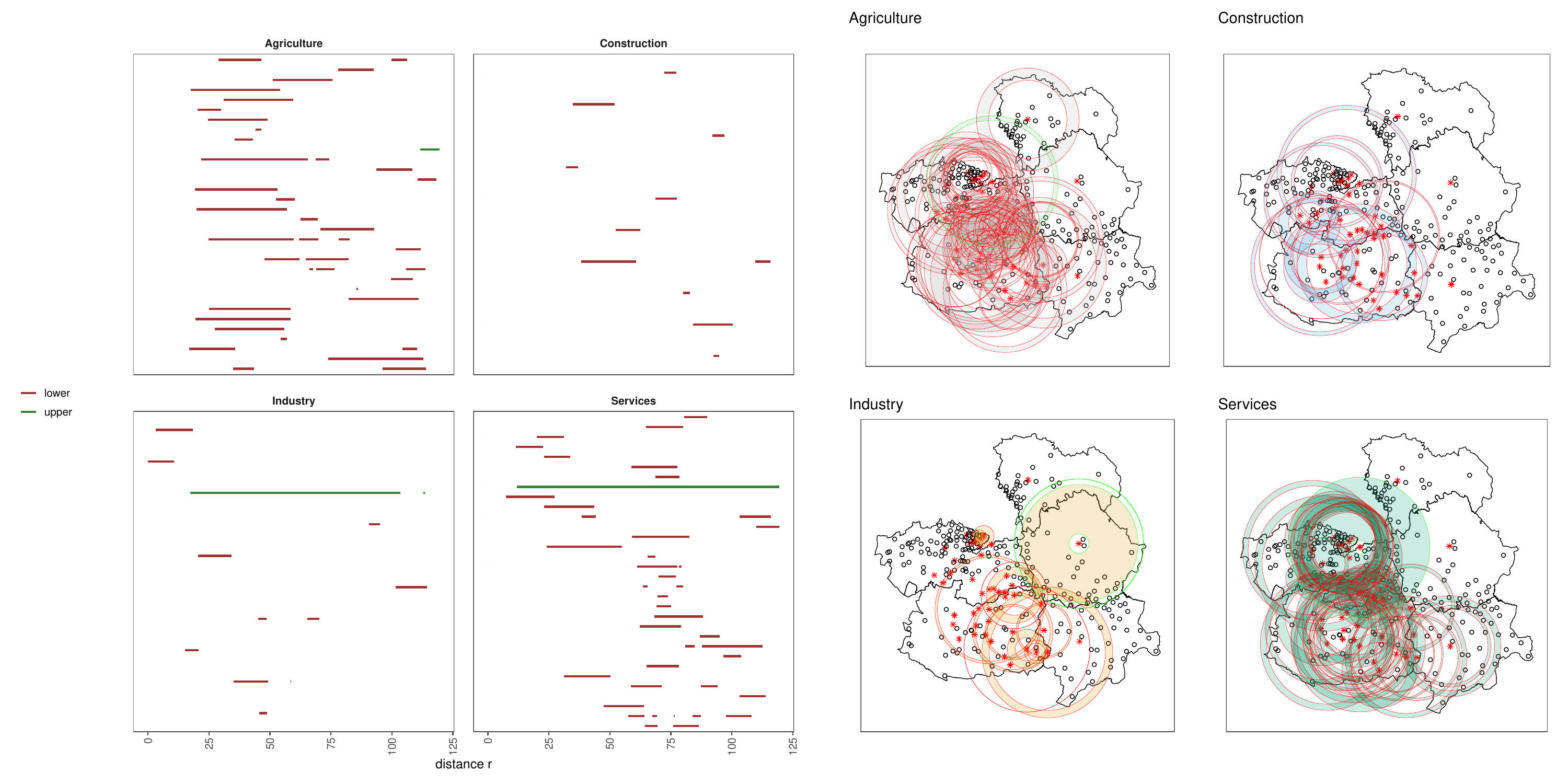}
    \caption{Significance bands obtained from employing $\gamma_i^{\text{clr},jj}$ in a global envelope test.}
    \label{fig:appendix_loctest_markvario}
\end{figure}
\end{landscape}

\begin{landscape}
\begin{figure}[h]
    \centering
    \includegraphics[width=\linewidth]{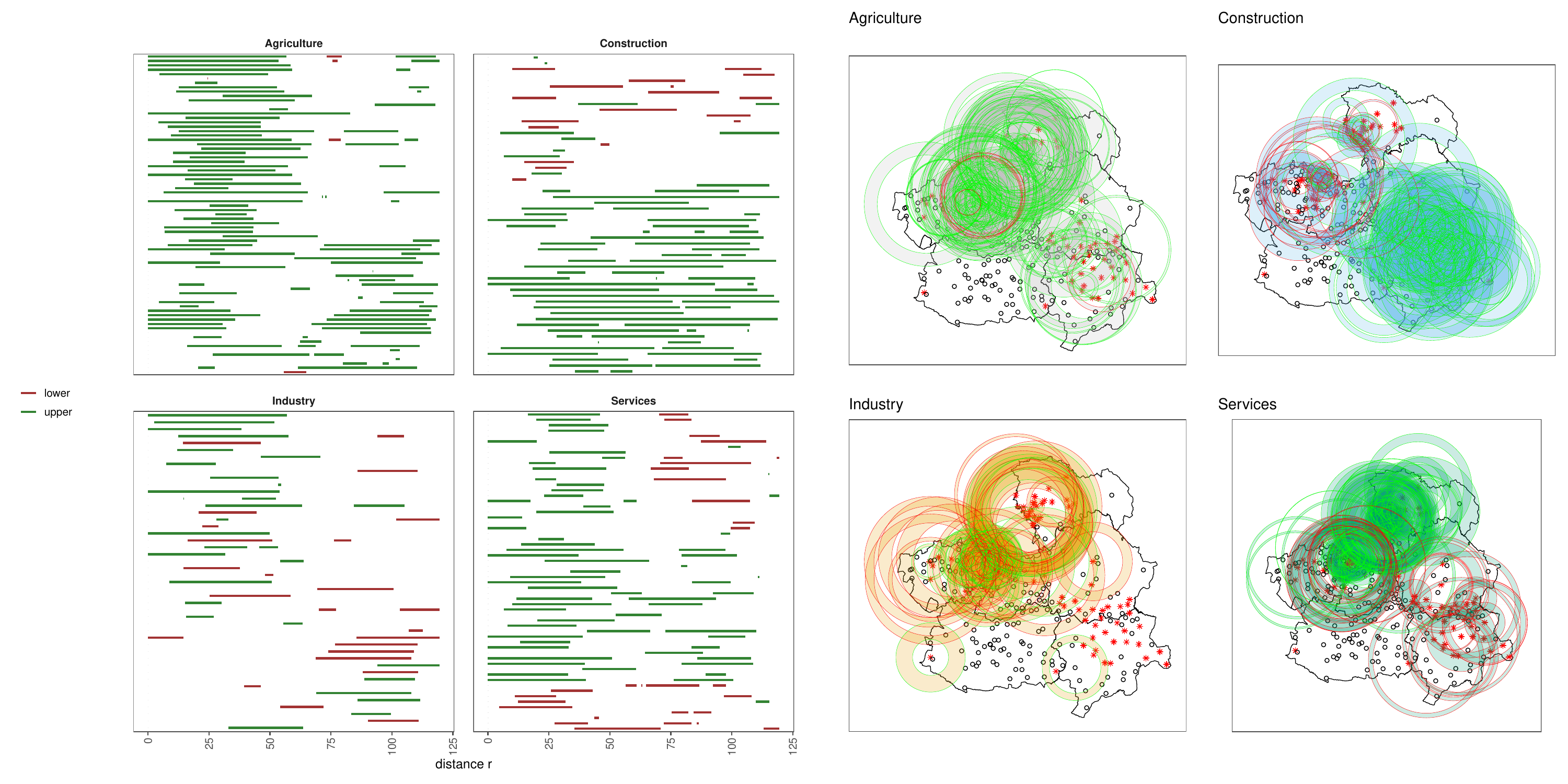}
    \caption{Significance bands obtained from employing $\iota_i^{\text{clr},jj}$ in a global envelope test.}
    \label{fig:appendix_loctest_shim}
\end{figure}
\end{landscape}

\printcredits

\bibliographystyle{cas-model2-names}

\bibliography{literature}

@book{aitchisonStatisticalAnalysisCompositional,
  title = {The {{Statistical Analysis}} of {{Compositional Data}}},
  author = {Aitchison, J.},
  year = 1986,
  series = {Monographs on {{Statistics}} and {{Applied Probability}}},
  publisher = {Springer Netherlands},
  address = {Dordrecht},
  doi = {10.1007/978-94-009-4109-0},
  langid = {english}
}

@article{anselinLocalIndicatorsSpatial,
  title = {Local {{Indicators}} of {{Spatial Association}} -- {{LISA}}},
  author = {Anselin, Luc},
  year = 1995,
  month = apr,
  journal = {Geographical Analysis},
  volume = {27},
  number = {2},
  pages = {93--115},
  issn = {0016-7363, 1538-4632},
  doi = {10.1111/j.1538-4632.1995.tb00338.x},
  urldate = {2025-01-20},
  langid = {english}
}

@misc{baddeleySpatstatSpatialPoint,
  title = {Spatstat: {{Spatial Point Pattern Analysis}}, {{Model-Fitting}}, {{Simulation}}, {{Tests}}},
  shorttitle = {Spatstat},
  author = {Baddeley, Adrian and Turner, Rolf and Rubak, Ege},
  year = {2025}
}

@Article{Eckardt:Moradi:currrent,
author={Eckardt, Matthias
and Moradi, Mehdi},
title={Marked Spatial Point Processes: Current State and Extensions to Point Processes on Linear Networks},
journal={Journal of Agricultural, Biological and Environmental Statistics},
year={2024},
volume={29},
number={2},
pages={346-378},
doi={10.1007/s13253-024-00605-1},
}

@article{Eckardt2023MultiFunctionMarks,
author = {Eckardt, Matthias and Comas, Carles and Mateu, Jorge},
title = {Summary Characteristics for Multivariate Function-Valued Spatial Point Process Attributes},
journal = {International Statistical Review},
volume = {93},
pages = {150–178},
year = {2025},
doi = {10.1111/insr.12582}
}

@misc{eckardt2024secondordercharacteristicsspatialpoint,
      title={Second-Order Characteristics for Spatial Point Processes with Graph-Valued Marks}, 
      author={Matthias Eckardt and Farnaz Ghorbanpour and Aila S{\"a}rkk{\"a}},
      year={2024},
      eprint={2410.16903},
      archivePrefix={arXiv},
      primaryClass={stat.ME},
      url={https://arxiv.org/abs/2410.16903}, 
}

@Article{Comas2011,
author={Comas, C.
and Delicado, P.
and Mateu, J.},
title={A Second Order Approach to Analyse Spatial Point Patterns with Functional Marks},
journal={Test},
year={2011},
volume={20},
number={3},
pages={503--523},
}

@article{diggleSpatialSpatioTemporalLogGaussian,
  title = {Spatial and {{Spatio-Temporal Log-Gaussian Cox Processes}}: {{Extending}} the {{Geostatistical Paradigm}}},
  shorttitle = {Spatial and {{Spatio-Temporal Log-Gaussian Cox Processes}}},
  author = {Diggle, Peter J. and Moraga, Paula and Rowlingson, Barry and Taylor, Benjamin M.},
  year = 2013,
  journal = {Statistical Science},
  volume = {28},
  number = {4},
  pages = {542--563},
  issn = {0883-4237},
  doi = {10.1214/13-STS441},
  urldate = {2025-02-15},
  langid = {english}
}

@article{eckardtLocalIndicatorsMark,
  title={Local Indicators of Mark Association for Marked Spatial Point Processes},
  author={Eckardt, Matthias and Moradi, Mehdi},
  journal={Journal of Computational and Graphical Statistics},
  number={just-accepted},
  pages={1--16},
  year={2026},
  publisher={Taylor \& Francis},
  doi = {10.1080/10618600.2026.2621093}
}

@article{eckardtSpatialPointProcesses,
  title = {On Spatial Point Processes with Composition-Valued Marks},
  author = {Eckardt, Matthias and Myllym{\"a}ki, Mari and Greven, Sonja},
journal = {International Statistical Review},
year = {2025},
doi = {https://doi.org/10.1111/insr.70019}
}

@book{illianStatisticalAnalysisModelling,
  title = {Statistical {{Analysis}} and {{Modelling}} of {{Spatial Point Patterns}}},
  author = {Illian, Janine and Penttinen, Antti and Stoyan, Helga and Stoyan, Dietrich},
  year = 2008,
  publisher = {Wiley}
}

@incollection{mateu-figuerasPrincipleWorkingCoordinates,
  title = {The {Principle} of {Working} on {Coordinates}},
  booktitle = {Compositional Data Analysis},
  author = {Mateu-Figueras, Glòria and Pawlowsky-Glahn, Vera and Egozcue, Juan José},
  year = {2011},
  pages = {29--42},
  publisher = {Wiley},
  doi = {10.1002/9781119976462.ch3},
  langid = {english}
}

@article{moradiInhomogeneousMarkCorrelation,
  title = {Inhomogeneous {{Mark Correlation Functions}} for {{General Marked Point Processes}}},
  author = {Moradi, Mehdi and Eckardt, Matthias},
  year = 2025,
  month = may,
  journal = {arXiv preprint arXiv.2505.24501},
  eprint = {2505.24501},
  primaryclass = {stat},
  urldate = {2025-06-04},
  archiveprefix = {arXiv},
  langid = {english}
}

@article{myllymakiGlobalEnvelopeTests,
  title = {Global {{Envelope Tests}} for {{Spatial Processes}}},
  author = {Myllym{\"a}ki, Mari and Mrkvi{\v c}ka, Tom{\'a}{\v s} and Grabarnik, Pavel and Seijo, Henri and Hahn, Ute},
  year = 2017,
  month = mar,
  journal = {Journal of the Royal Statistical Society Series B: Statistical Methodology},
  volume = {79},
  number = {2},
  pages = {381--404},
  issn = {1369-7412, 1467-9868},
  doi = {10.1111/rssb.12172},
  urldate = {2025-03-18},
  langid = {english}
}

@article{ripleyModellingSpatialPatterns,
  title = {Modelling {{Spatial Patterns}}},
  author = {Ripley, B. D.},
  year = 1977,
  month = jan,
  journal = {Journal of the Royal Statistical Society Series B: Statistical Methodology},
  volume = {39},
  number = {2},
  pages = {172--192},
  issn = {1369-7412, 1467-9868},
  doi = {10.1111/j.2517-6161.1977.tb01615.x},
  urldate = {2025-05-07},
  langid = {english}
}

@article{schlatherDetectingDependenceMarks,
  title = {Detecting {{Dependence Between Marks}} and {{Locations}} of {{Marked Point Processes}}},
  author = {Schlather, Martin and Ribeiro, Paulo J. and Diggle, Peter J.},
  year = 2004,
  journal = {Journal of the Royal Statistical Society. Series B (Statistical Methodology)},
  volume = {66},
  number = {1},
  pages = {79--93},
  issn = {1369-7412},
  urldate = {2025-04-30}
}

@article{shimataniPointProcessesFineScale,
  title = {Point {{Processes}} for {{Fine-Scale Spatial Genetics}} and {{Molecular Ecology}}},
  author = {Shimatani, Kenichiro},
  year = 2002,
  month = apr,
  journal = {Biometrical Journal},
  volume = {44},
  number = {3},
  pages = {325--352},
  issn = {03233847, 15214036},
  doi = {10.1002/1521-4036(200204)44:3<325::AID-BIMJ325>3.0.CO;2-B},
  urldate = {2025-06-04},
  langid = {english}
}



\end{document}